\documentclass[prl,twocolumn,showpacs,superscriptaddress]{revtex4}

\usepackage{epsfig,graphicx,amssymb}

\begin{document}

\title{Field squeeze operators in optical cavities with atomic ensembles}

\author{R. Guzm\'an}

\affiliation{Departamento de F\'{\i}sica, Universidad de Santiago
de Chile, Casilla 307, Correo 2, Santiago, Chile}

\author{J.~C.~Retamal}

\affiliation{Departamento de F\'{\i}sica, Universidad de Santiago
de Chile, Casilla 307, Correo 2, Santiago, Chile}

\author{E.~Solano}

\affiliation{Max-Planck-Institut f{\"u}r Quantenoptik,
Hans-Kopfermann-Strasse 1, D-85748 Garching, Germany}

\affiliation{Secci\'{o}n F\'{\i}sica, Departamento de Ciencias,
Pontificia Universidad Cat\'{o}lica del Per\'{u}, Apartado 1761,
Lima, Peru}

\author{N.~Zagury}

\affiliation{Instituto de F\'{\i}sica, Universidade Federal do Rio
de Janeiro, Caixa Postal 68528, 21945-970 Rio de Janeiro, Brazil}

\date{\today}

\begin{abstract}
We propose a method of generating \textit{unitarily} single and two-mode
field squeezing in an optical cavity with an atomic cloud. Through a
suitable laser system, we are able to engineer a squeeze field operator
\textit{decoupled} from the atomic degrees of freedom, yielding a large
squeeze parameter that is scaled up by the number of atoms, and realizing
degenerate and non-degenerate parametric amplification. By means of the
input-output theory we show that ideal squeezed states and perfect squeezing
could be approached at the output. The scheme is robust to decoherence
processes.
\end{abstract}


\maketitle

Squeezing can be defined, in a harmonic oscillator, as the
reduction of quantum fluctuations in a certain quadrature below
the vacuum level, at the expense of increasing them in its
canonically conjugate variable~\cite{MandelWolf}. The possibility
of manipulating quantum fluctuations was first noticed by Caves
\textit{et al.}~\cite{Caves}, with the aim of precision
measurements. Since then, much effort has been devoted to it
through theoretical proposals and experimental
implementations~\cite{Wineland}. Recently, with the advent of
quantum information and communication, entangled squeezed states
of the electromagnetic field~\cite{Lam} have led to the
realization of continuous variable teleportation~\cite{Kimble}.
Also, by improving the yet low squeeze parameters, it is expected
that two-mode squeezed states will lead to efficient distribution
of entanglement and implementation of quantum
channels~\cite{ClarkKraus}. Two-mode (polarization) squeezing has
already been realized by means of Kerr nonlinearity in optical
fibers~\cite{Lorenz} and with cold atomic clouds in optical
cavities~\cite{Giacobino}. Recently, theoretical and experimental
developments relating atomic ensembles and quantum information
devices, like entanglement~\cite{Polzik} and exchange of
information between light and atomic states~\cite{Hammerer}, have
raised justified expectations on related topics. However, to our
knowledge, an effective and tunable field squeeze
operator~\cite{MandelWolf} has not yet been proposed or realized.

In this letter, we present a method that produces \textit{single
and two-mode squeeze operators}, decoupled from the atomic degrees
of freedom, acting on a cavity containing an atomic ensemble. The
squeeze parameters scale up with the interaction time and with the
number of atoms present in the interaction region of the cavity.
This method shows to be robust against decoherence processes, like
spontaneous emission, and does not require strong coupling regime
or strong atomic localization. Furthermore, we use the
input-output formalism to show that it is possible to generate
two-photon coherent states~\cite{Yuen} (or ideal squeezed
states~\cite{idealCaves}) and to approach perfect squeezing at the
output field, allowing the study of their features in a wide range
of parameters.

Our model consists of an ensemble of $N$ identical three-level
atoms inside an optical cavity~\cite{Molmer,Giacobino}. For the
sake of generality, we assume that the atoms occupy random
positions $\mathbf{r}_{k}$ ($k=1,...,N$) along the spatial
distribution of two cavity modes, $u_{a}(\mathbf{r})$ and
$u_{b}(\mathbf{r})$. Each atom interacts with these two quantized
modes and with two properly tuned lasers, as sketched in Fig. 1,
yielding a couple of independent Raman laser systems. The
associated Hamiltonian can be written as
\begin{eqnarray}
H=H_{\mathrm{0}}+H_{\mathrm{int}} ,
\end{eqnarray}
with
\begin{equation}
H_{\mathrm{0}}=\hbar \omega _{a}a^{\dag }a+\hbar \omega _{b}b^{\dag }b+\hbar
\sum\limits_{k=1}^{N}\sum\limits_{i=0}^{2}\omega _{i}\left| i\right\rangle
_{k}\left\langle i\right| ,
\end{equation}
and
\begin{eqnarray}  \label{hint}
H_{\mathrm{int}} &=&\sum\limits_{k=1}^{N}\left\{ \left( \Omega
_{k1}\left| 0\right\rangle _{k}\left\langle 1\right| e^{-i\nu _{1}
t}+\Omega _{k1}^{\ast }\left| 1\right\rangle_{k} \left\langle
0\right| e^{i\nu _{1}t}\right) \right.
\nonumber \\
&&+\left( \Omega _{k2}\left| 0\right\rangle _{k}\left\langle
2\right| e^{-i\nu _{2} t}+\Omega _{k2}^{\ast }\left|
2\right\rangle_{k} \left\langle
0\right| e^{i\nu _{2} t}\right)  \nonumber \\
&&+\left( g_{ka}\left| 0\right\rangle _{k}\left\langle 2\right|
a+g_{ka}^{\ast }\left| 2\right\rangle _{k}\left\langle 0\right| a^{\dag
}\right)  \nonumber \\
&&+\left( g_{kb}\left| 0\right\rangle _{k}\left\langle 1\right|
b+g_{kb}^{\ast }\left| 1\right\rangle _{k}\left\langle 0\right|
b^{\dag }\right) \} .
\end{eqnarray}
Here, $a$ ($a^{\dagger }$) and $b$ ($b^{\dagger }$) are the
annihilation (creation) operators associated with two cavity
modes, with frequencies $\omega _{a}$ and $\omega _{b}$,
respectively. Atomic states $|i\rangle $ ($i=0,1,2$) have Bohr
frequencies $\omega _{i}$ and are coupled in two simultaneous
Lambda configurations~\cite{LawEberly}. Atomic transitions
$|1\rangle \leftrightarrow |0\rangle $ and $|2\rangle
\leftrightarrow |0\rangle $ are coupled through classical fields
with coupling constants $\Omega _{k1} =
\Omega_{1}v_{1}(\mathbf{r}_{k})$ and $\Omega _{k2}=\Omega_{2} \,
v_{2}( \mathbf{r}_{k})$, and also through the two cavity modes,
$b$ and $a$, with coupling constants
$g_{kb}=g_{b}\,u_{b}(\mathbf{r}_{k}) $ and
$g_{ka}=g_{a}\,u_{a}(\mathbf{r}_{k})$, respectively.

In the interaction picture, the associated Hamiltonian reads
\begin{eqnarray}
H_{\mathrm{I}} & \!\! = \!\! & \hbar \sum\limits_{k=1}^{N}\left\{
\Omega _{k1}\left| 0\right\rangle _{k}\left\langle 1\right|
e^{-i\Delta _{1}t}+\Omega _{k1}^{\ast }\left| 1\right\rangle_{k}
\left\langle 0\right| e^{i\Delta _{1}t}\right.  \nonumber \\
&&+\Omega _{k2}\left| 0\right\rangle _{k}\left\langle 2\right|
e^{-i\Delta _{2}t}+\Omega _{k2}^{\ast }\left| 2\right\rangle_{k}
\left\langle 0\right| e^{i\Delta _{2}t}  \nonumber \\
&&+g_{ka}\left| 0\right\rangle _{k}\left\langle 2\right|
ae^{-i\tilde{\Delta} _{1}t}+g_{ka}^{\ast }\left| 2\right\rangle
_{k}\left\langle 0\right| a^{\dag }e^{i\tilde{\Delta}_{1}t}
\nonumber \\ &&\left. +g_{kb}\left| 0\right\rangle
_{k}\left\langle 1\right| be^{-i\tilde{ \Delta}_{2}t}+g_{kb}^{\ast
}\left| 1\right\rangle _{k}\left\langle 0\right| b^{\dag
}e^{i\tilde{\Delta}_{2}t}\right\} ,
\end{eqnarray}
where $\Delta _{i}=\omega _{i}-\omega _{0}+\nu _{i}$,
$\tilde{\Delta} _{i}=\Delta _{i}-\delta _{i}$ ($i=1,2$), and also
$\tilde{\Delta}_{1}=\omega _{2}-\omega _{0}+\omega _{a}$,
$\tilde{\Delta}_{2}=\omega _{1}-\omega _{0}+\omega _{b}$. We
consider dispersive detunings $\{ |(\Delta _{i}-\Delta
_{j})|,|(\tilde{\Delta}_{i}-\tilde{\Delta}_{j})|,|(\tilde{\Delta}
_{i}-\Delta _{j})|,|\Delta_{i}|,|\tilde{\Delta}_{i}| \} \gg
\{|\delta _{i}|,|g_{ka}|,|g_{kb}|,|\Omega _{k1}|,|\Omega
_{k2}|\}$, with $i=1,2$ ($ i\neq j$), $k=1,...,N$. Then, we
eliminate adiabatically level $ |0\rangle $ and obtain the
effective Hamiltonian
\begin{eqnarray}
\!\!\!\!\!\!H_{\mathrm{II}} &\!\!\!=\!\!\!&\hbar \sum_{k}
\bigg\lbrack \left( \frac{|g_{kb}|^{2}}{\tilde{\Delta}_{2}}
b^{\dag} b + \frac{|\Omega _{k1}|^{2}}{\Delta _{1}}\right) \left|
1 \right\rangle _{k}\left\langle 1 \right|  \nonumber
\label{firsteffectiveHamiltonian} \\ && + \left(
\frac{|g_{ka}|^{2}}{\tilde{\Delta}_{1}} a^{\dag} a + \frac{|\Omega
_{k2}|^{2}}{\Delta _{2}}\right) \left| 2 \right\rangle
_{k}\left\langle 2 \right|  \nonumber \\ && + \left( \frac{\Omega
_{k1}g_{ka}^{\ast }}{\Delta _{1}}a^{\dag }e^{-i\delta
_{1}t}+\frac{g_{kb}\Omega _{k2}^{\ast }}{\Delta _{2}} be^{i\delta
_{2}t}\right) \sigma _{k}^{-}\,  \nonumber \\ && + \left(
\frac{\Omega _{k2}g_{kb}^{\ast }}{\Delta _{2}}b^{\dag }e^{-i\delta
_{2}t}+\frac{g_{ka}\Omega _{k1}^{\ast }}{\Delta _{1}} ae^{i\delta
_{1}t}\right) \sigma _{k}^{+} \bigg\rbrack ,
\end{eqnarray}
where $\sigma _{k}^{+}=\left| 1\right\rangle _{k}\left\langle 2\right| $ and
$\sigma _{k}^{-}=\sigma _{k}^{+\dagger }=\left| 2\right\rangle
_{k}\left\langle 1\right| $ are raising and lowering atomic operators,
respectively. For simplicity, we have discarded terms that require an
initial population of level $|0\rangle $.

By making the unitary transformation $e^{i A t / \hbar}$, with
\begin{equation}
A= \hbar \widetilde{\delta }(a^{\dag }a+b^{\dag }b) + \hbar
\sum_{k} (  \frac{|\Omega _{k1}|^{2}}{\Delta _{1}}\left\vert
1\right\rangle _{k}\left\langle 1\right\vert + \frac{ |\Omega
_{k2}|^{2}}{\Delta _{2}}\left\vert 2\right\rangle _{k}\left\langle
2\right\vert ) ,
\end{equation}
where $\ \widetilde{\delta }=(\delta _{2}+\delta _{1})/2$ , we obtain the
new Hamiltonian
\begin{eqnarray}
H_{\mathrm{III}} &=&\hbar \{-\tilde{\delta}(a^{\dag }a+b^{\dag
}b)+ \nonumber \\ && \sum_{k}(
\frac{|g_{ka}|^{2}}{\tilde{\Delta}_{1}} a^{\dag } a
 \left\vert 2 \right\rangle _{k} \left\langle 2 \right\vert +
 \frac{ |g_{kb}|^{2}}{ \tilde{\Delta}_{2}} b^{\dag } b \left\vert
1\right\rangle _{k}\left\langle 1\right\vert  \nonumber \\
&&+[(\tilde{\Omega}_{ka}a+\tilde{\Omega}_{kb}^{\ast }b^{\dag
})e^{-i\delta _{k}t}\sigma _{k}^{+}+\mathrm{H.c.}])\} .
\end{eqnarray}
Here, $\delta _{k}=(\delta _{2}-\delta _{1})/2+|\Omega
_{k2}|^{2}/\Delta _{2}-|\Omega _{k1}|^{2}/\Delta _{1}$, and
$\tilde{\Omega}_{ka}=\Omega _{k1}^{\ast }g_{ka}/\Delta _{1},$
$\tilde{\Omega}_{kb}=\Omega _{k2}^{\ast }g_{kb}/\Delta _{2}$.

\begin{figure}[t]
\begin{center}
\includegraphics[width=0.3\textwidth]{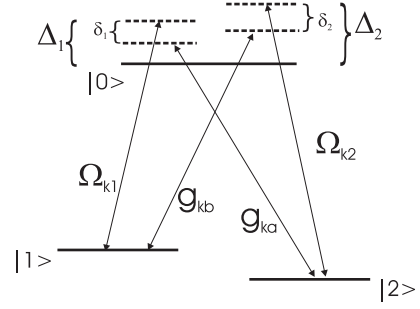}
\end{center}
\par
\vspace*{-0.5cm} \caption{Each three-level atom is driven with two
classical fields, with frequencies $\protect\nu _{1}$ and
$\protect\nu _{2}$, establishing a couple of Raman laser systems
through two cavity modes.}
\end{figure}

We make now the unitary transformation $e^{iV(t) / \hbar}$, with
$V(t)=-i \hbar \sum_{k}\sigma _{k}^{-}\left(
\tilde{\Omega}_{ka}^{\ast }a^{\dag }+ \tilde{\Omega} _{kb}b\right)
(e^{i\delta _{k}t}-1)/\delta _{k}+\mathrm{H.c.}$ , assuming
$|\delta _{k}|t\gg 1$, $|\delta _{k}|\gg \{|\tilde{\Omega}_{ka}|,|
\tilde{ \Omega}_{kb}| , \frac{|g_{ka}|^{2}}{\tilde{\Delta}_{1}} ,
\frac{ |g_{kb}|^{2}}{ \tilde{\Delta}_{2}} , \tilde{\delta}\}$, and
obtain
\begin{eqnarray}  \label{HIV}
& \!\!\!\!\! H_{\mathrm{IV}} & \!\!\! = \! \hbar
\{-\tilde{\delta}(a^{\dag }a \! + \! b^{\dag }b) + \!\! \sum_{k}(
\frac{|\tilde{\Omega}_{kb}|^{2}}{ \delta _{k}} | 2 \rangle_k
\langle 2 | \! - \! \frac{|\tilde{\Omega} _{ka}|^{2}}{\delta _{k}}
| 1 \rangle_k \langle 1 |  \nonumber \\ && + \frac{|g_{ka}
|^{2}}{\tilde{\Delta}_{1}}a^{\dag }a\left\vert 2\right\rangle
_{k}\left\langle 2\right\vert + \frac{|g_{kb}|^{2}}{\tilde{
\Delta}_{2}} b^{\dag }b \left\vert 1\right\rangle _{k}\left\langle
1\right\vert  \nonumber \\ && - \sigma
_{k}^{z}\,[\,\frac{|\tilde{\Omega}_{ka}|^{2}}{\delta _{k}} a^{\dag
}a + \frac{|\tilde{\Omega}_{kb}|^{2}}{\delta _{k}}b^{\dag }b
\nonumber \\ && + \frac{\tilde{\Omega}_{ka}^{\ast
}\tilde{\Omega}_{kb}^{\ast }}{\delta _{k}}a^{\dag }b^{\dag
}+\frac{\tilde{\Omega}_{ka}\tilde{\Omega}_{kb}}{\delta _{k}} ba ]
) \} ,
\end{eqnarray}
where $\sigma _{k}^{z}=|1\rangle _{k}\langle 1|-|2\rangle _{k}\langle 2|$.
We assume that all atoms are initially in the ground state, $\Pi
_{k}|2\rangle _{k}$, which allows us to replace $\sigma _{k}^{z}$ by $-1$ in
Eq.~(\ref{HIV}). We require the condition
\begin{equation}
\tilde{\delta}=\sum_{k}( \frac{|\tilde{\Omega}_{ka}|^{2}}{\delta
_{k}} + \frac{|g_{ka}|^{2}}{\tilde{\Delta}_{1}})=\sum_{k}
\frac{|\tilde{\Omega} _{kb}|^{2}}{\delta _{k}} , \label{condition}
\end{equation}
which can be easily satisfied by adjusting properly the classical
field strengths $\Omega _{k1}/\Omega _{k2}$ and the ratio $\Delta
_{1}/\Delta _{2}$ . Then, up to a constant term, $H_{\mathrm{IV}}$
can be rewritten as
\begin{equation}
H_{\mathrm{IV}}=\hbar (\Omega a^{\dag }b^{\dag }+\Omega ^{\ast }ba),
\label{FinalHamiltonian}
\end{equation}
where $\Omega =\sum_{k}\tilde{\Omega}_{ka}^{\ast
}\tilde{\Omega}_{kb}^{\ast }/\delta _{k}.$ Therefore, the time
evolution operator in the Schr\"{o}dinger picture reads
\begin{equation}
U(\tau )\simeq e^{-i\tilde{H}_{0}\tau /\hbar }U^{\mathrm{ND}}(\tau ),
\end{equation}
where we made $e^{iV(t) / \hbar}\simeq 1$, consistently with
approximations made before, and defined $\tilde{H}_{0}$ $=H_{0} +
A$ and
\begin{equation}
U^{\mathrm{ND}}=e^{(\xi ^{\ast }ab-\xi \,a^{\dagger }b^{\dagger })}.
\label{UND}
\end{equation}
Here, explicitly, $\xi =\tau \sum_{k}\Omega _{k1}\Omega
_{k2}g_{ka}^{\ast }g_{kb}^{\ast }/(\delta _{k}\Delta _{1}\Delta
_{2})$ is a squeeze parameter that scales with the number of atoms
$N$, $\tau $ being the interaction time. The time evolution
operator in Eq.~(\ref{UND}) is a unitary two-mode squeeze operator
that is decoupled from the atomic degrees of freedom, producing
two-mode squeezing on any initial field state. In particular,
given that at room temperature an optical cavity field is in the
vacuum state, a two-mode squeezed vacuum will be naturally
produced. Eq.~(\ref{UND}) corresponds to a physical implementation
of a non-degenerate (ND) parametric oscillator in the domain of
cavity QED and atomic clouds. Implementation of a degenerate (D)
parametric oscillator is straightforward if we consider mode $b$
identical to mode $a$, yielding $U^{\mathrm{D}}=e^{(\xi ^{\ast
}a^{2}-\xi \,a^{\dagger 2})}$.

At this point, we will make some experimental considerations,
stressing that each physical implementation will require specific
adaptations. In fact, to assure a large squeezing in a fixed
quadrature of the cavity mode, all atoms in the interaction volume
should contribute coherently. Let us consider an optical cavity
with cylindrical symmetry around the z-axis with $u_{a}(
\mathbf{r}_{k})\sim \sin (q_{a}z_{k})f_{a}(\rho _{k})e^{im\varphi
_{k}} $, $ u_{b}(\mathbf{r}_{k})\sim \sin (q_{b}z_{k})f_{b}(\rho
_{k})e^{-im\varphi _{k}}$. We choose the classical fields to
counterpropagate perpendicular to the axis of the cavity, so as to
warrant a coherent atomic contribution in the effective
interaction volume when $d|q_{a}-q_{b}|\ll 1$ and $w|\nu _{1}-\nu
_{2}|/c\ll 1$, $d$ being the beam widths and $w\sim d$ the waist
of the modes. These conditions relax the typical requirement of
atomic localization inside a wavelength, and can be easily
satisfied, in general, if the two lower levels are separated by a
small splitting compared to optical frequencies.

We consider a low density vapor of $^{85}\mathrm{Rb}$ in an
optical cavity. The two lower levels $|1\rangle $ and $|2\rangle$
are the ground state $(5S_{1/2})$ hyperfine levels ($F=2$ and
$F=3$), separated by $3 $GHz, while level $|0\rangle $ is the
first excited state $(5P_{3/2})$, yielding optical transitions of
$780\mathrm{nm}$. Using the cavity parameters of
Ref.~\cite{RubidiumRempe}, we have $w\sim 35\mu \mathrm{m}$,
homogeneous laser beams of width $d\sim 50\mu \mathrm{m}$, and an
interaction volume of $\sim 10^{-7}\mathrm{cm^{3}}$. We choose,
for example, $\Omega _{k1}$ $\sim 100g$, $\Omega _{k2} = \Omega
_{k1} / 10$, $\Delta _{2} = 2 \Delta _{1}$, $g_{a} \sim g_{b} =
g$, and dispersive condition $\Delta _{1} / \Omega _{k1} = 100$.
These values and Eq.~(\ref{condition}) are enough to estimate all
relevant parameters, while satisfying strictly all requirements to
derive $H_{\rm IV}$ of Eq.~(\ref{FinalHamiltonian}). In
particular, we calculate $\delta_k = -(1-1/400) g$, $\delta_1 =
-3g / 400$, $\delta_2 = -g / 80$, $\tilde{\delta} = - g / 100$, an
effective coupling $\Omega = -g / 5$, and $N = 4 \times 10^4$,
corresponding to a density of $\lesssim 10^{12}/\mathrm{cm^{3}}$
(small enough to prevent coherence losses due to collisions). This
is just a rather conservative set of parameters, for a chosen
experimental setup~\cite{RubidiumRempe}, from a wide range of
possibilities.

The maximal value of the squeeze parameter in Eq.~(\ref{UND}), for
the same example, is $|\xi|_{max} \sim (| g | / 5) \tau_{\rm
diss}$, with $\tau_{\mathrm{diss}}\equiv 1 / \kappa {\bar{n}}$,
$\kappa$ being the cavity decay rate and ${\bar{n}}$ the mean
number of cavity photons. Given that for squeezed vacuum
$\bar{n}=\sinh ^{2}(2|\xi |)$, and with a conservative $| g | /
\kappa \sim 5$, the present scheme should be able to produce field
squeezing $\sim 70 \%$, which is a competitive value when compared
with recent achievements~\cite{Giacobino}. However, as we will see
in the second part of the manuscript, the condition $| \Omega | =
| g | / 5 = \kappa$ is enough to approach, theoretically, perfect
squeezing at the cavity output.

The noisy effect of spontaneous emission will be negligible here,
for typical values of individual atomic emission rate $\Gamma$ and
in presence of a large number of atoms. It is possible to estimate
that even for a high squeeze parameter $\xi$, very few photons,
$N_{\Gamma } \sim N\Gamma (\Omega _{k}/\Delta )^{2} \tau
_{\mathrm{ diss}} \ll N$, would be spontaneously emitted from the
whole cloud. For the realistic parameters of our previous example,
$ N_{\Gamma }<1\ll N \sim 4 \times 10^4$.

Now, we will concentrate on the squeezing properties of the
outgoing cavity field. We recall that the output field that has
been considered for diverse applications and can be measured
through standard optical procedures~\cite {Giacobino}.

We consider the input-output theory, successfully applied to the
study of the parametric amplifier~\cite{GardinerZoller}, for the
case of two cavity modes driven by the effective nonlinear
interaction in Eq.~(\ref{UND}) and by external (axial) laser
fields. The classical fields drive cavity modes $a$ and $b$ with
strengths $\epsilon _{a}$ and $\epsilon _{b}$, respectively. We
assume that each cavity mode interacts with an independent heat
bath such that, in the Markov approximation, the following coupled
Langevin equations are produced
\begin{eqnarray}
\dot{a} &=&-i\epsilon _{a}^{\ast }+\frac{\Omega }{2}\,b^{\dag }-\frac{\kappa
_{a}}{2}a-\sqrt{\kappa _{a}}\,c_{\mathrm{in}}(t)  \nonumber \\
\dot{b} &=&-i\epsilon _{b}^{\ast }+\frac{\Omega }{2}\,a^{\dag }-\frac{\kappa
_{b}}{2}b-\sqrt{\kappa _{b}}\,d_{\mathrm{in}}(t).  \label{uno}
\end{eqnarray}
Here, $c_{\mathrm{in}}(t)$ and $d_{\mathrm{in}}(t)$ are
annihilation operators associated with the input fields, $\kappa
_{a}$ and $\kappa _{b}$ are the cavity decay rates of modes $a$
and $b$, and we have considered $ \Omega \rightarrow i \Omega /2$
($\Omega$ real) to match standard notation~\cite {GardinerZoller}.
Then, Eqs.~(\ref{uno}) can be rewritten as
\begin{eqnarray}
\dot{a^{\prime }} &=&\frac{\Omega }{2}\,{b^{\prime }}^{\dag }-\frac{\kappa
_{a}}{2}{a^{\prime }}-\sqrt{\kappa _{a}}\,c_{\mathrm{in}}(t)  \nonumber \\
\dot{b^{\prime }} &=&\frac{\Omega }{2}\,{a^{\prime }}^{\dag }-\frac{\kappa
_{b}}{2}b^{\prime }-\sqrt{\kappa _{b}}\,d_{\mathrm{in}}(t),  \label{unox}
\end{eqnarray}
where the transformations
\begin{eqnarray}
a &=&a^{\prime }+\alpha _{0}  \nonumber \\
b &=&b^{\prime }+\beta _{0},  \label{transf}
\end{eqnarray}
with $\alpha _{0}=2i(\kappa _{b}\epsilon _{a}^{\ast }-\Omega \epsilon
_{b})/(\Omega ^{2}-\kappa _{a}\kappa _{b})$ and $\beta _{0}=2i(\kappa
_{a}\epsilon _{b}^{\ast }-\Omega \epsilon _{a})/(\Omega ^{2}-\kappa
_{a}\kappa _{b})$, have been realized.

For the sake of convenience, we calculate the solutions of Eq.
(\ref{unox}) in frequency domain
\begin{eqnarray}
\tilde{a^{\prime}}(\omega ) &=&\frac{2\sqrt{\kappa _{a}} \, \beta }{ \Omega
^{2} - \alpha \beta }{\tilde c}_{\mathrm{in}}(\omega )+\frac{2\sqrt{\kappa
_{b}} \, \Omega }{\Omega ^{2} - \alpha \beta }{\tilde d}_{\mathrm{in}}^{\dag
}(-\omega )  \nonumber \\
\tilde{b^{\prime}}(\omega ) &=&\frac{2\sqrt{\kappa _{b}} \, \alpha }{ \Omega
^{2} - \alpha \beta }{\tilde d}_{\mathrm{in}}(\omega )+\frac{2\sqrt{\kappa
_{a}} \, \Omega }{ \Omega ^{2} - \alpha \beta }{\tilde c}_{\mathrm{in}
}^{\dag }(-\omega ) ,  \label{soluinside}
\end{eqnarray}
where ${\tilde x} ( \omega )$ is the Fourier transform of each operator $x
(t)$, and $\alpha =\kappa _{a}-2i\omega $ and $\beta =\kappa _{b}-2i\omega$.

Following a standard procedure, and undoing the transformation of
Eq.~(\ref {transf}), the output fields can be determined as a
function of the input fields~\cite{GardinerZoller} ,
\begin{eqnarray}
c_{\mathrm{out}}(\omega )= & - & \sqrt{\kappa_a} \alpha_0 \delta ( \omega )
\nonumber \\
&+& \frac{\Omega ^{2} + \alpha ^{\ast } \beta}{\Omega ^{2}- \alpha \beta }
c_{\mathrm{in}} \left( \omega \right) +\frac{2\Omega \sqrt{%
\kappa_{a}\kappa_{b}}}{\Omega ^{2} - \alpha \beta } d_{\mathrm{in}}^{\dag
}\left( -\omega \right) ,  \nonumber \\
d_{\mathrm{out}}(\omega ) = & - & \sqrt{\kappa_b} \beta_0 \delta ( \omega )
\nonumber \\
&+& \frac{\Omega ^{2}+\alpha \beta ^{\ast }}{\Omega ^{2} - \alpha
\beta } d_{ \mathrm{in}}\left( \omega \right) +\frac{2\Omega
\sqrt{\kappa_{a}\kappa_{b}} }{\Omega ^{2} - \alpha \beta }
c_{\mathrm{in}}^{\dag }\left( -\omega \right)
.  \nonumber \\
&&  \label{outputs1}
\end{eqnarray}

As suggested in \cite{Loudon,CavesSchumaker}, two-mode field
quadratures can be defined as $X=\left( a+b+a^{\dag }+b^{\dag
}\right) /2^{3/2}$ and $ Y=-i\left( a+b-a^{\dag }-b^{\dag }\right)
/2^{3/2}$. From the solutions in Eq. (\ref{outputs1}), we can
calculate, at resonance,
\begin{equation}
(\Delta X_{\mathrm{out}})^2 = \frac{1}{4} \left( \frac{ \Omega + \sqrt{
\kappa_a \kappa_b } }{ \Omega - \sqrt{ \kappa_a \kappa_b } } \right) ^2
\label{xquadrature}
\end{equation}
\begin{equation}
(\Delta Y_{\mathrm{out}})^2 = \frac{1}{4} \left( \frac{ \Omega - \sqrt{
\kappa_a \kappa_b } }{ \Omega + \sqrt{ \kappa_a \kappa_b } } \right) ^2 ,
\label{yquadrature}
\end{equation}
where $(\Delta X_{\mathrm{out}})^2 = \langle X_{\mathrm{out}}^2
\rangle - \langle X_{\mathrm{out}} \rangle ^2$, and $(\Delta
Y_{\mathrm{out}})^2 = \langle Y_{\mathrm{out}}^2 \rangle - \langle
Y_{\mathrm{out}} \rangle ^2$. Note that $(\Delta
X_{\mathrm{out}})^2 (\Delta Y_{\mathrm{out}})^2 = 1 / 16$ , like
it should be for a minimum uncertainty field state. If the
nonlinear coupling vanishes, then $(\Delta X_{\mathrm{out}})^2 =
(\Delta Y_{\mathrm{\ out }})^2 = 1 / 4$, as it should be for a
coherent state (including the particular case of the vacuum
state). However, in general, Eqs.~(\ref {xquadrature}) and
(\ref{yquadrature}) show that $1/4 \leq (\Delta X_{
\mathrm{out}})^2 < \infty$ and $0 \leq (\Delta Y_{\mathrm{out}})^2
\leq 1 / 4 $. The reduction parameter $r \equiv 2 | \xi |$,
assuming $(\Delta Y_{ \mathrm{out}} )^2 = e^{-r}/4$, is

\begin{eqnarray}
r = - 2 \ln \left| \frac{\Omega - \sqrt{\kappa_a \kappa_b}}{\Omega + \sqrt{
\kappa_a \kappa_b}} \right| .  \label{squeeze}
\end{eqnarray}

We have shown, in principle, that quadrature $Y_{\mathrm{out}}$ at
the output can achieve perfect squeezing $(\Delta
Y_{\mathrm{out}})^2 = 0$, when $\Omega = \sqrt{\kappa_a \kappa_b}$
($r \rightarrow \infty$), at the expense of large fluctuations in
$X_{\mathrm{out}}$. Clearly, feedback and saturation effects will
prevent perfect squeezing from happening but those considerations
are beyond the scope of this work. This limiting situation, known
for the degenerate case, is still valid for nondegenerate two-mode
squeezing in presence of classical drivings at the input. Note
that, even if the fluctuations in Eqs.~(\ref{xquadrature}) and
(\ref{yquadrature}) do not depend on the driving parameters
$\epsilon_a$ and $\epsilon_b$, these yield effective displacements
at the output, see Eq. (\ref{outputs1}), with amplitudes
\begin{eqnarray}
\alpha_{\mathrm{eff}} = \sqrt{\kappa_a} \alpha_0 = \frac{2 i \sqrt{\kappa_a}
( \kappa_b \epsilon_a^{*} - \Omega \epsilon_b ) } {\Omega^2 - \kappa_a
\kappa_b }  \nonumber \\
\beta_{\mathrm{eff}} = \sqrt{\kappa_b} \beta_0 = \frac{2 i \sqrt{\kappa_b} (
\kappa_a \epsilon_b^{*} - \Omega \epsilon_a ) } {\Omega^2 - \kappa_a
\kappa_b } \,\, .  \label{disp}
\end{eqnarray}

This fact suggests that the output field could be interpreted
either as a two-mode two-photon coherent state~\cite{Yuen}, $S
(\xi_{\mathrm{eff}} )
D(\alpha_{\mathrm{eff}})D(\beta_{\mathrm{eff}}) | 0 \rangle$,
where $| \xi_{ \mathrm{eff}} | = r / 2$ is given by
Eq.~(\ref{squeeze}) and $\alpha_{ \mathrm{eff}}$ and
$\beta_{\mathrm{eff}}$ by Eqs.~(\ref{disp}), or, equivalently, as
an ideal two-mode squeezed state~\cite{idealCaves}, $
D(\alpha^{\prime}_{\mathrm{eff} })D(\beta^{\prime}_{\mathrm{eff}})
S ( \xi^{\prime}_{\mathrm{eff}} ) | 0 \rangle$. Note that the
experimentally tuned parameters $\alpha_{\mathrm{eff}}$ and
$\beta_{\mathrm{\ eff}}$ diverge under the condition of perfect
squeezing, $\Omega = \sqrt{ \kappa_a \kappa_b}$, and they vanish
for $\kappa_b \epsilon_a^{*} = \Omega \epsilon_b$ , $\kappa_a
\epsilon_b^{*} = \Omega \epsilon_a$. These cases do not violate
energy conservation and are consistent with the model.

In conclusion, we have presented a method to implement effectively
and efficiently single mode and two-mode field squeeze operators.
This is realized through a suitable laser system acting on an
atomic cloud inside a cavity, implementing degenerate and
non-degenerate parametric amplification in a novel manner. The
collective action of the atoms in the cloud yields enhancement of
the squeeze parameter that is proportional to the number of atoms
and the interaction time. This unitary procedure squeezes any
field state and in particular the initial vacuum field. By means
of the input-output theory, we have shown that it is possible to
generate conditions for approaching perfect squeezing and ideal
squeezed states at the cavity output in a controlled manner.
Extensions to the case of ring cavities are straightforward and
may simplify the requirements of the present proposal.
Experimental achievement of these goals should contribute to the
study of fundamental aspects in quantum noise reduction, and to
the implementation of diverse quantum communication schemes, like
entanglement distribution and remote exchange of quantum
information.

R.G. and J.C.R. acknowledge support from Grants No. Fondecyt
1030189, No. Milenio ICM P02-049, and No. MECESUP USA0108. E.S. is
grateful for the hospitality at USACH (Santiago de Chile), at UFRJ
(Rio de Janeiro), and acknowledges support from EU project RESQ.
N.Z. acknowledges support from CNPq, FAPERJ, and thanks Claudio
Lenz Cesar for helpful discussions, and is grateful for the
hospitality at USACH.

\end{document}